\documentclass[preprint,12pt, a4paper]{elsarticle}

\usepackage{amssymb}
\usepackage[hyphens,spaces,obeyspaces]{url}
\usepackage[colorlinks=true,allcolors=blue]{hyperref}

\usepackage{float}

\usepackage{xcolor}

\restylefloat{table}

\journal{SoftwareX}

\begin{document}

\begin{frontmatter}



\title{PyPartMC: A Pythonic interface to a particle-resolved, Monte Carlo aerosol simulation framework}


\author[ATMS]{Zachary~D'Aquino}
\author[AGH]{Sylwester~Arabas}
\author[ATMS]{Jeffrey~H.~Curtis}
\author[ECE]{Akshunna~Vaishnav}
\author[ATMS]{Nicole~Riemer}
\author[MECHSE]{Matthew~West}

\address[ATMS]{Atmospheric Sciences, University of Illinois Urbana-Champaign, Urbana, IL, USA}
\address[AGH]{Physics and Applied Computer Science, AGH Univeristy of Krakow, Kraków, Poland}
\address[ECE]{Electrical and Computer Engineering, University of Illinois Urbana-Champaign, Urbana, IL, USA}
\address[MECHSE]{Mechanical Science and Engineering, University of Illinois Urbana-Champaign, Urbana, IL, USA}

\begin{abstract}

PyPartMC is a Pythonic interface to PartMC, a stochastic, particle-resolved aerosol model implemented in Fortran. Both PyPartMC and PartMC are free, libre, and open-source. PyPartMC reduces the number of steps and mitigates the effort necessary to install and utilize the resources of PartMC. Without PyPartMC, setting up PartMC requires: working with UNIX shell, providing Fortran and C libraries, and performing standard Fortran and C source code configuration, compilation and linking. This can be challenging for those less experienced with computational research or those intending to use PartMC in environments where provision of UNIX tools is less straightforward (e.g., on Windows). PyPartMC offers a single-step installation/upgrade process of PartMC and all dependencies through the pip Python package manager on Linux, macOS, and Windows. This allows streamlined access to the unmodified and versioned Fortran internals of the PartMC codebase from both Python and other interoperable environments (e.g., Julia through PyCall). In particular, PyPartMC can be set up to handle the time-stepping loop for PartMC simulations making it possible to couple PartMC with other Python-interoperable packages, for either online diagnostics or additional simulation logic. Altogether, users of PyPartMC can setup, run, process and visualize output of PartMC simulations using a single general-purpose programming language.
\end{abstract}

        
\end{frontmatter}

\section*{Required Metadata}

\section*{Current code version}

\begin{table}[H]\footnotesize
\begin{tabular}{|l|p{5.9cm}|p{5.9cm}|}
\hline
\textbf{Nr.} & \textbf{Code metadata description} & \\
\hline
C1 & Current code version & 1.0.0 (planned for release with the paper) \\
\hline
C2 & Permanent link to code/repository used for this code version & \url{https://github.com/open-atmos/PyPartMC} \\
\hline
C3 & Code Ocean compute capsule & None \\
\hline
C4 & Legal Code License & GNU GPL v3.0 \\
\hline
C5 & Code versioning system used & git \\
\hline
C6 & Software code languages, tools, and services used & C++, Fortran, C, Python, CMake, Jupyter \\
\hline
C7 & Compilation requirements, operating environments \& dependencies & Dependencies: SuiteSparse, SUNDIALS, CAMP, PartMC, pybind11-JSON, pybind11, JSON-Fortran, nlohmann/json, netCDF 
\newline OS: Linux, macOS, Windows\\
\hline
C8 & Link to developer documentation/manual & \url{https://open-atmos.github.io/PyPartMC/} \\
\hline
C9 & Support email for questions & please use GitHub issue tracker \\
\hline
\end{tabular}
\caption{Code metadata}
\label{} 
\end{table}

\section{Motivation and significance}

Open-source software fosters collaboration and accountability within the scientific community through joint development and maintenance ventures that accelerate scientific discovery and progress \citep{heistermann2015, bangerth2013, piva2012, kogut2001}. The interoperability and accessibility of proven computational tools produced through open-source projects can benefit from the development of bindings to popular high-level, general-purpose programming languages, such as Python. This helps in~providing researchers with a community of contributors and users to~support individual research pursuits. In this paper, we focus on the value of open-source software and the provision of a robust Python interface to an established Fortran codebase. The PyPartMC project documented in this paper has specific applicability in the field of aerosol science; however, the outlined design applies in general to the development of Pythonic interfaces to existing open-source models implemented in Fortran.

PyPartMC is a free, libre, and open-source Pythonic interface to the Fortran-implemented stochastic, particle-resolved aerosol model PartMC \citep{Riemer_2009}, currently at version 2.7.0. PartMC simulates the microphysical processes that aerosol particles undergo during their lifetimes in the atmosphere. These process include new particle formation, emission from primary sources, Brownian diffusion, aqueous chemical processing, and removal by dry deposition and nucleation scavenging. PartMC categorically tracks the chemical composition of each computational particle, allowing the user to comprehensively simulate the mixing state and other physiochemical properties of an aerosol population. PartMC typically runs in conjunction with a solver that simulates the dynamic partitioning of semi-volatile organic and inorganic gases.

PartMC is a powerful open-source tool for simulating the dynamics of aerosol populations; however, running a PartMC simulation requires knowledge of shell and the use of a netCDF-Fortran library, which can present a~significant hurdle for those with less experience in computation or those intending to use PartMC on platforms less compliant with the standard UNIX development environment. Current installation (or upgrade) of PartMC requires working command of Fortran compilers, CMake, netCDF, and awareness of the exact directories for a number of libraries on the user's machine. This process can be tedious and prone to errors due to the number of steps involved. 

A primary motivating factor for developing PyPartMC is to remove limitations to installation and setup of PartMC dependent on the individual user's computational background. PyPartMC offers streamlined use of PartMC by providing a single-step installation process and flexibility in selecting the libraries and frameworks, for example, Jupyter and Dask \citep{dask} from the Python ecosystem, each individual user wishes to apply cooperatively during independent execution of simulations and routines based on PartMC. Prior to the development of PyPartMC, auxiliary programs written in Fortran were necessary to process the output from a PartMC simulation. In addition to providing a simpler installation experience, PyPartMC saves users of PartMC from having to code anything in Fortran. The interface capitalizes on many Python functionalities, such as garbage collection and native data types. Overall, one of the aims of the project is to enable researchers using PartMC to perform all tasks pertaining to setting up, running, post-processing and analysing simulations using a single programming language. 

PyPartMC is available as source and pre-compiled packages on \url{pypi.org} and can be installed at the command line via  \verb|pip install PyPartMC|. 
In~case pre-compiled binaries for a given hardware or software configuration are unavailable, the \verb|pip install| command orchestrates compilation of the package. The continuous integration (CI) setup maintained on the GitHub platform verifies a consistent installation experience across Linux, macOS and Windows platforms and is used for building pre-compiled packages. For~Linux builds, standarized `manylinux` environments are used to ensure compatibility with a wide range of systems. The source package ships with all the dependencies included (maintained using git submodules what streamlines updates and ensures version traceability). The binary packages have all dependencies statically linked, further minimizing installation environment constraints. The three-language (Fortran, C, C++) build automation is handled through CMake and wired to the Python package installation, making the process invisible to the user. Test automation is achieved with pytest \citep{pytest}. 

Python's stature as a high-level, truly general-purpose programming language with its outstanding readability and ease of learning appeals to a much larger group of programmers that exceeds the size of the more specialized Fortran community. Exposing the PartMC codebase to the Python ecosystem enables individuals to take advantage of the expansive community of Python developers and users that provide support through online documentation and forums. These resources are not as prolific in the smaller Fortran community. In general, a lower-level language, like Fortran, requires more lines of code to accomplish the same task in a higher-level language. A programmer's productivity is constrained by the number of lines of code needed to accomplish a given task, and it has been shown that programmers write roughly the same number of line of code per unit time independent of the language used~\citep{Wilson_2014}. Consequently, tasks can be accomplished in shorter time with less code using higher-level programming languages, like Python. PyPartMC assists with comprehensibility and allows users to build code to their personal specifications at the rate of a high-level language with the performance boost of a low-level language.

All dependencies, such as SuiteSparse \citep{suitesparse}, SUNDIALS \citep{sundials}, Chemistry Across Multiple Phases (CAMP) \citep{camp}, netCDF, and nlohmann/json \citep{nlohmann_json}, are free, open-source, and maintained with versioned releases, constituting a practical tool set for reproducible computational research. Simplifying installation and setup of PartMC through a Pythonic interface allows PyPartMC to be used to support the development of robust aerosol model benchmarking architectures. The ability to leverage resources from the PartMC simulation suite while taking advantage of existing libraries and frameworks that are ubiquitous in the scientific community provides the necessary infrastructure for PyPartMC to participate in the movement to improve STEM education through active learning applications \citep{Freeman_2014}. One of the technologies that facilitates active learning by streamlining dissemination of interactive computational worksheets is Jupyter notebooks, which PyPartMC provides for usage examples (see Section~\ref{sec:notebooks} below). Noteworthy, the employment of Jupyter notebooks for active learning applications has already been demonstrated in the very field of aerosol science \citep{petters2021}.

\section{Software description}

\subsection{Particle-resolved modeling with PartMC}
PartMC (Particle Monte Carlo) is a stochastic, particle-resolved aerosol box model, which resolves the composition of many individual aerosol particles within a well-mixed volume of air \citep{Riemer_2009, DeVille2011, Curtis2016, DeVille2019}. The documentation for PartMC can be found at \url{http://lagrange.mechse.illinois.edu/partmc/partmc-2.6.1/doc/html/index.html}. Aerosol processes such as coagulation, nucleation, emission, and deposition are implemented in a Monte Carlo fashion, i.e., by sampling the population with appropriate probabilities. Our particle-resolved approach uses a large number of discrete computational particles ($10^4$ to $10^6$) to represent the particle population of interest. Each particle is represented by a ``composition vector'', which stores the mass of each constituent species within each particle and evolves over the course of a simulation according to various chemical or physical processes. The ``weighted flow algorithm'' \citep{DeVille2011, DeVille2019} improves the model efficiency and reduces ensemble variance by assigning each computational particle a number concentration that this particle corresponds to. In contrast to other particle-based methods used for example in cloud physics,  PartMC does not track the position of the computational particle in physical space. To account for multi-phase chemical processes, PartMC needs to be coupled to a chemistry code, e.g. the Model for Simulating Aerosol Interactions and Chemistry (MOSAIC) \citep{zaveri2008} or CAMP \citep{camp}. PartMC has been used in many applications as a box model, but has also been coupled to the Weather Research and Forecast model (WRF) to simulate aerosol processes and transport in a 3D model domain, with WRF as the host model supplying the flow field. Since the particles' locations in 3D space are not stored, we use a stochastic sampling approach to move particles between grid boxes~\citep{Curtis2017}.

\subsection{Software architecture}
\begin{figure}
    \centering
    \includegraphics{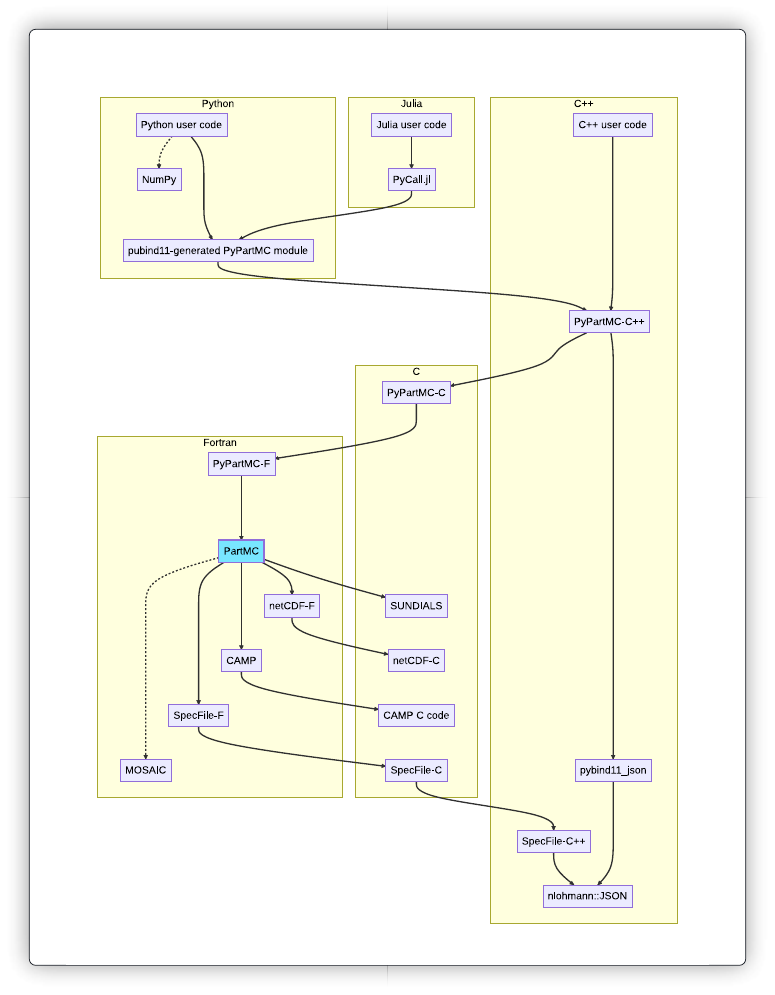}
    \caption{Schematic of architecture of PyPartMC depicting the coupling between Fortran, C and C++ internals and the user code written in Python or Julia.}
    \label{fig:architecture}
\end{figure}

PyPartMC is written in Fortran, C, and C++. Fortran wrappers of the existing PartMC Fortran code are defined in the PyPartMC codebase using the standard Fortran \verb|ISO_C_BINDING| functionalities. These are used in concert with a thin C-implemented wrapping layer to define C++ bindings to PartMC Fortran datatypes and routines. The pybind11 framework is then used to build the Python API from these C++ bindings (see Fig~\ref{fig:architecture} for an overview of the design). 
PartMC has been implemented in an object-oriented manner, leveraging Fortran derived types which are extensively used throughout the modular PartMC codebase. These derived types are exposed in the PyPartMC bindings as Python classes.
Unlike the case of black-box Python bindings to Fortran code \citep[e.g., as in][]{vandenOord_2020}, PyPartMC reveals the underlying internals of the Fortran modules to Python users.

PyPartMC is a Pythonic interface that will be natural to use for those familiar with the Python programming language but retains PartMC jargon. Python garbage collection deallocates Fortran objects instantiated through PyPartMC, ensuring user code is succinct and memory-safe. Instantiation and deserialization of exposed PartMC structures are handled using JSON-like data structures which provides an authentic Pythonic experience. Users leverage Python's built-in data types, syntax, and comprehension to construct and store simulation parameters which eliminates the need for any input text files used in typical PartMC simulation executions.

PyPartMC has continuous integration workflows for Linux, macOS, and Windows handled through GitHub Actions which verifies that any new feature added to the project is checked through automated testing and compiles successfully on all targeted platforms. Unit tests account for a significant portion of the codebase and certify that developments do not cause regressions by changing the behavior of previously working code. Test execution is orchestrated by pytest, and the PyPartMC unit test suite assures information and execution pass as expected between the wrapping code and PartMC internals. PartMC contains its own testing suite that checks the underlying physical processes and their representations within the model.

Docstring-based API documentation is published on the web upon each pull request merge (https://open-atmos.github.io/PyPartMC/). New versions of PyPartMC are easily disseminated thanks to automated uploads to \url{pypi.org} triggered by GitHub releases. 

\subsection{Software functionality}
This software architecture preserves the performance advantages of a Fortran-implemented codebase and allows PyPartMC to simultaneously serve as a C++ API to the PartMC Fortran internals. This structure also circumvents the need for any temporary files and does not require any additional routines to be added to the PartMC codebase for interfacing. 

The Python packaging infrastructure expedites dissemination of both source and binary packages ensuring code version traceability for PyPartMC bindings, PartMC itself, and all other dependencies. Noteworthy, this---together with the employed static linkage of all dependencies---effectively circumvents the notorious difficulties in disseminating binary version of multi-dependency Fortran software rooted in the lack of binary compatibility between compilers, and even between versions of the same Fortran compiler~\citep{Kedward_et_al_2022}.

The design also capitalizes on Python's ability to integrate with other languages, so PyPartMC can be called from alternative environments. 
Similarly, PyPartMC can be used to couple PartMC with simulation or diagnostics components implemented in other software and hardware (e.g., GPU) solutions interoperable with Python. To this end, PyPartMC provides the option to perform time-stepping within PartMC simulations in Python, and this is one of the rationales for designing the package as an API to PartMC internals rather than merely wrapping PartMC solver control procedures. One of these environments includes the Julia programming language which has been adopted by many in the field of computational research, including aerosol scientists \citep{petters2021}. An example employing PyPartMC from Julia (using PyCall) is maintained in the PyPartMC README file on GitHub and shown in Section~\ref{sec:basic_examples} below (also see diagram in Fig.~\ref{fig:architecture}). 

The design choice of building an API for PartMC rather than using shell wrappers eliminates overhead and allows for more freedom in external coupling logic, providing the added benefit of supporting adoption of new features added to PartMC with compiler static analysis of the wrapping code. In cases where flags are read exclusively from PartMC input text files, no new API elements are needed. PyPartMC can simply be compiled against another version of PartMC to allow access to newly added flags in the PartMC codebase; however, this may not be possible if the interface were constructed in another manner. Additionally, prior installation of PartMC is not required, static linkage of PartMC and all dependencies ensures that the versions of the wrapper, library and dependencies match.

PyPartMC currently offers access to a subset of PartMC functions (approximately 20\%). While the process of giving access to a larger subset is still ongoing, not all functions will need to be given access to. The choice of making functions available has been largely user-driven; starting from the most fundamental functions (returning particle masses), we added more recently functions that are related to specific post-processing functionalities (such as computing critical supersaturation and aerosol mixing state parameters).

Finally, regarding the performance of the implemented solution, it is worth highlighting that the overhead of starting the Python interpreter, loading the package and even (optionally) performing time-stepping loop in Python is only noticeable in simulations with impractically small state-vector sizes.
Testing setups featuring $10^2$ and $10^4$ particles, this constant overhead was measured to amount to roughly 10\% and 1\% of the process CPU time, respectively.

\section{Basic examples in Python and Julia}\label{sec:basic_examples}

\begin{figure*}
\begin{tabular}{p{.5\textwidth}p{.5\textwidth}}

  {\scriptsize\bf a: Python code (with embedded data)}\newline
  \includegraphics[width=.5\textwidth]{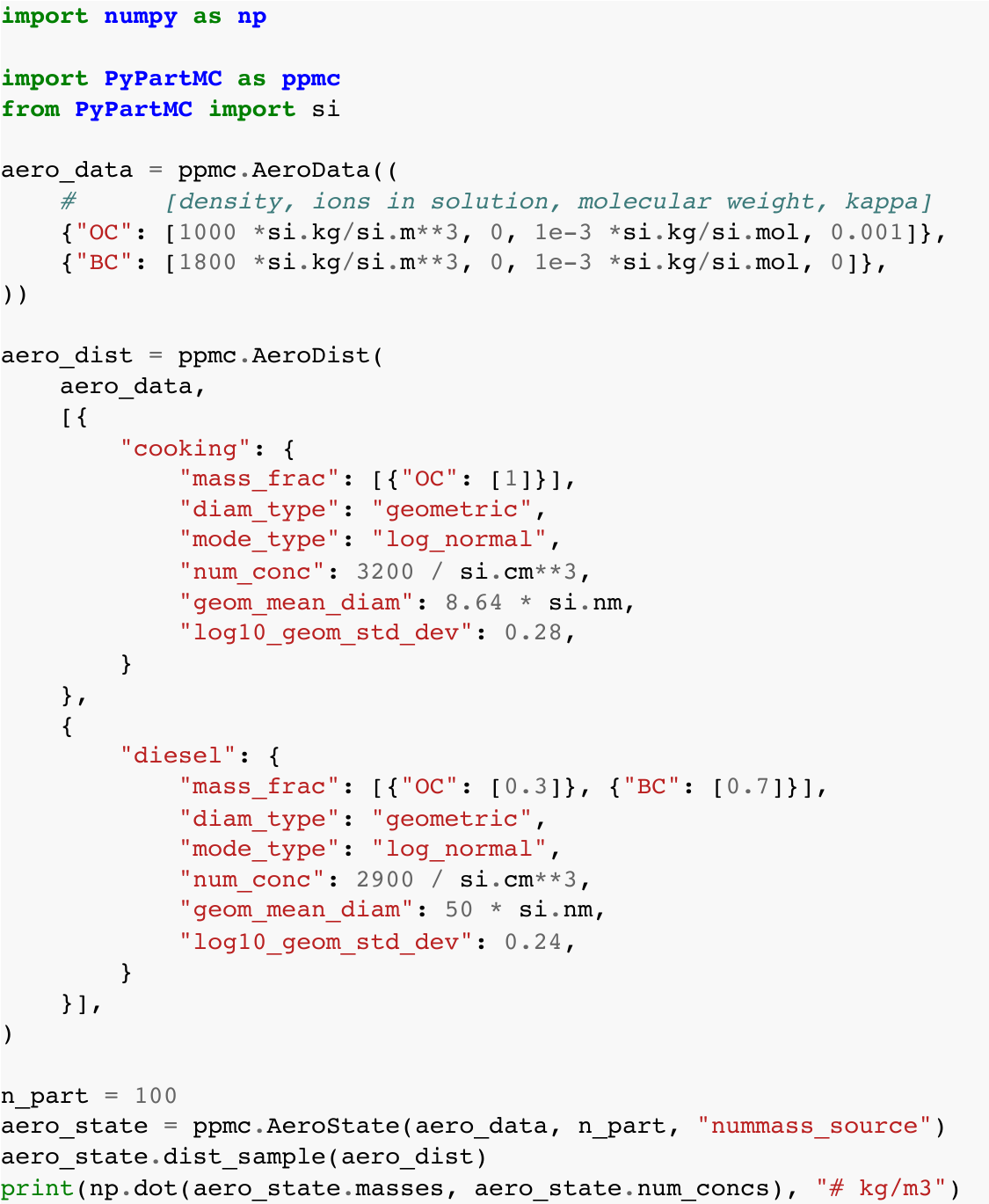}
  \newline

  {\scriptsize\bf b: Julia code (with embedded data)}\newline
  \includegraphics[width=.5\textwidth]{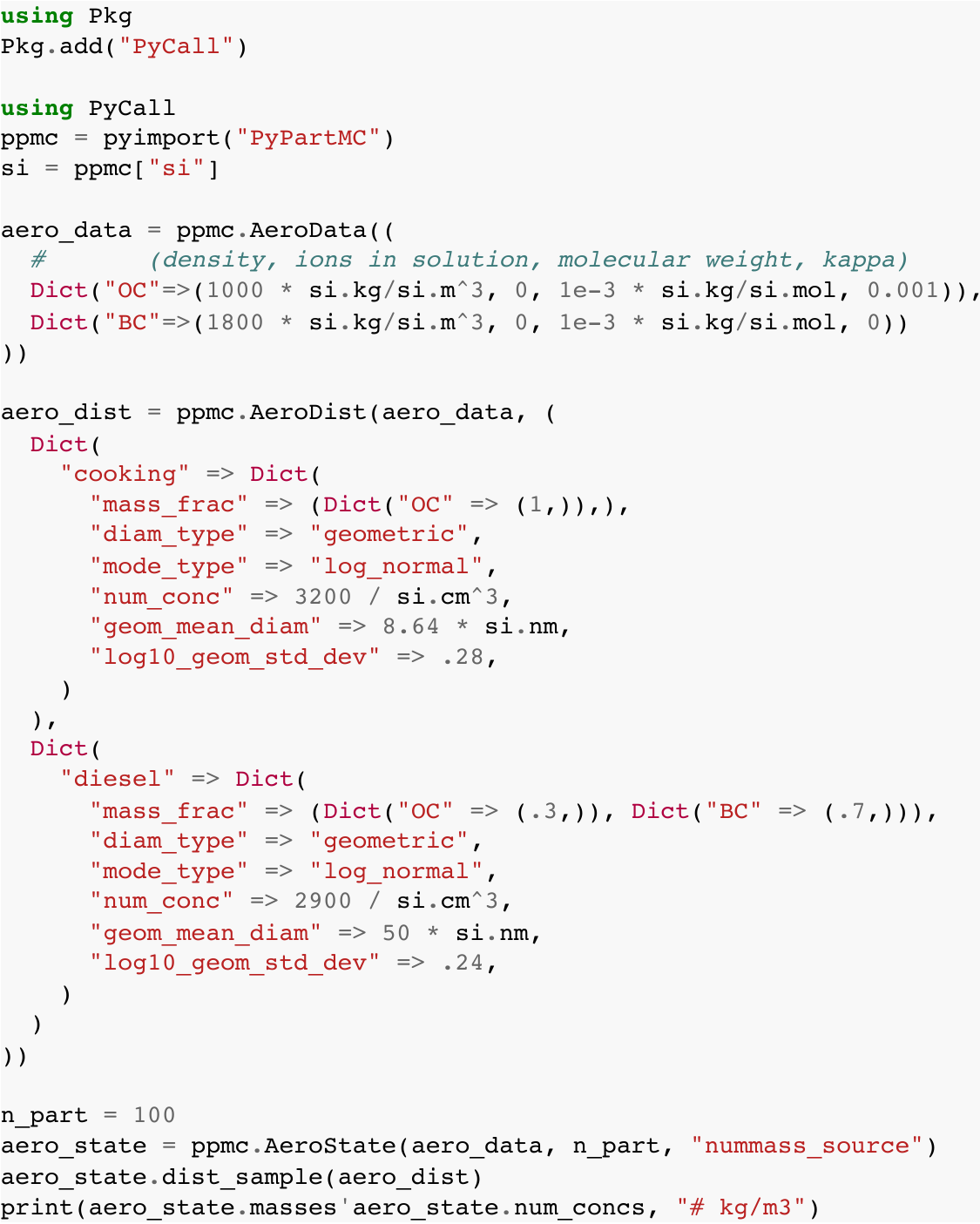}
  
  &
  
  {\scriptsize\bf c: Fortran code}\newline
  \includegraphics[width=.5\textwidth]{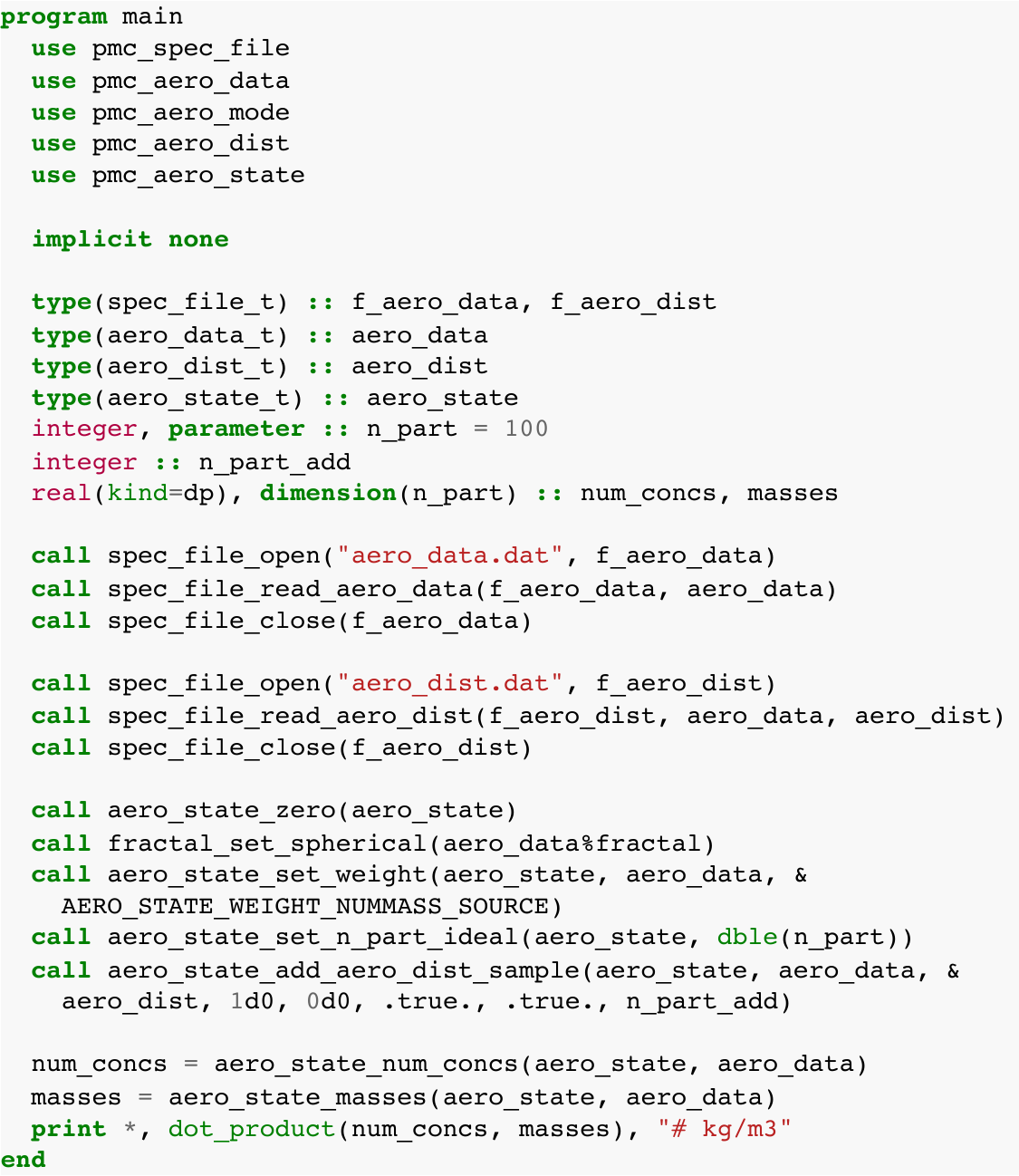}
  \newline
  
  {\scriptsize\bf d: aero\_dist.dat file (for Fortran code)}\newline
  \includegraphics[width=.5\textwidth]{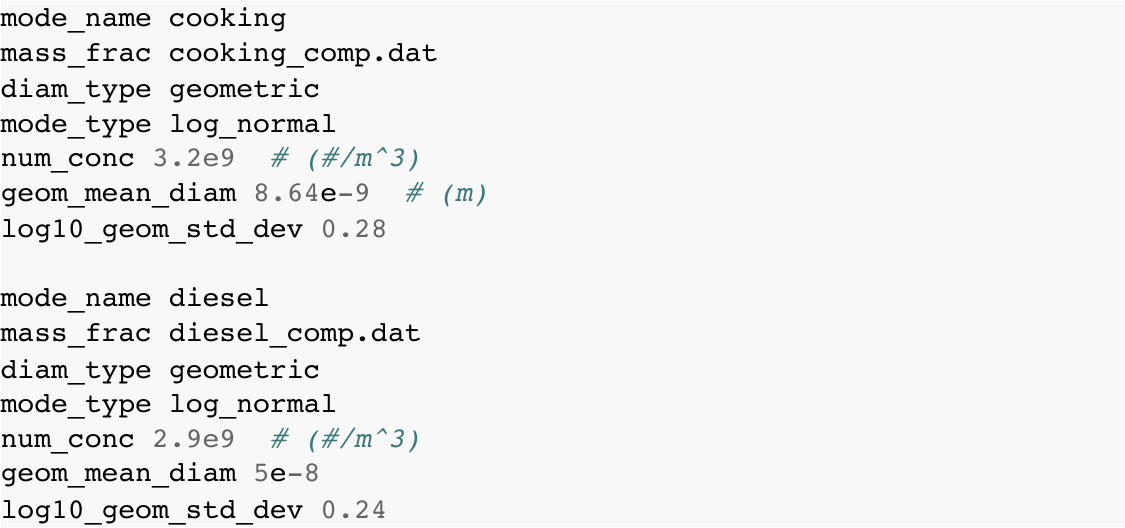}
  \newline

  {\scriptsize\bf e: cooking\_comp.dat file (for Fortran code)}\newline
  \includegraphics[width=.5\textwidth]{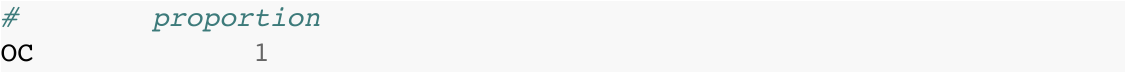}
  \newline
  
  {\scriptsize\bf f: diesel\_comp.dat file (for Fortran code)}\newline
  \includegraphics[width=.5\textwidth]{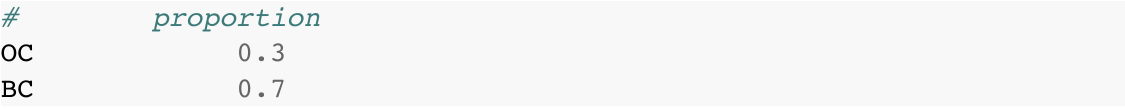}
  
\end{tabular}
\vspace{-1.5em}
\caption{Code listings depicting basic usage of PartMC data structures and algorithms in: (a) Python through PyPartMC; (b) Julia through PyCall and PyPartMC; (c-f) Fortran.}
\label{fig:basic_examples}
\end{figure*}

Figure \ref{fig:basic_examples} depicts how the identical task of randomly sampling particles from
  an aerosol size distribution in PartMC can be done in three different 
  programming languages.
Listing (a) presents Python code employing PyPartMC.
Listing (b) is an analogous code written in Julia where PyPartMC is 
  accessed using PyCall to fully interoperate with Python.
Listings (c-f) depict how this task is achieved using Fortran
  native to PartMC. In this case, the parameters of the particle size distribution are provided in external text files.

\section{Illustrative examples (Jupyter notebooks)}\label{sec:notebooks}

PyPartMC ships with a collection of Jupyter notebook examples with Python kernel displaying how PyPartMC can be leveraged to access PartMC resources in a Python environment. In the README of the PyPartMC repository, users can navigate to a collection of badges corresponding to each example that will render the notebook with nbviewer, execute in the cloud via Binder or Google Colab \citep{google_colab}, or create a standalone web application with Voilà. These example notebooks work out of the box and constitute experiential and manageable tools users can employ to explore the utility of the API and immediately begin building on the provided code. The notebooks illustrate how results obtained through PyPartMC based on PartMC simulations can be contained in a web-based notebook and easily distributed with minimal software and hardware dependencies. Example~1 showcases the comparison of two different software packages that both calculate
the water uptake of aerosol particles. Example~2 illustrates the simulation of a typical PartMC scenario and post-
processing of the raw data to obtain aggregate quantities that are common in aerosol science. Example~3 illustrates the simulation of
PartMC to supply particle populations to an external Python package to compute aerosol optical properties.

\subsection{Example \#1: Comparison with PySDM}
\begin{figure}[ht]
\centering
\includegraphics[width=\textwidth]{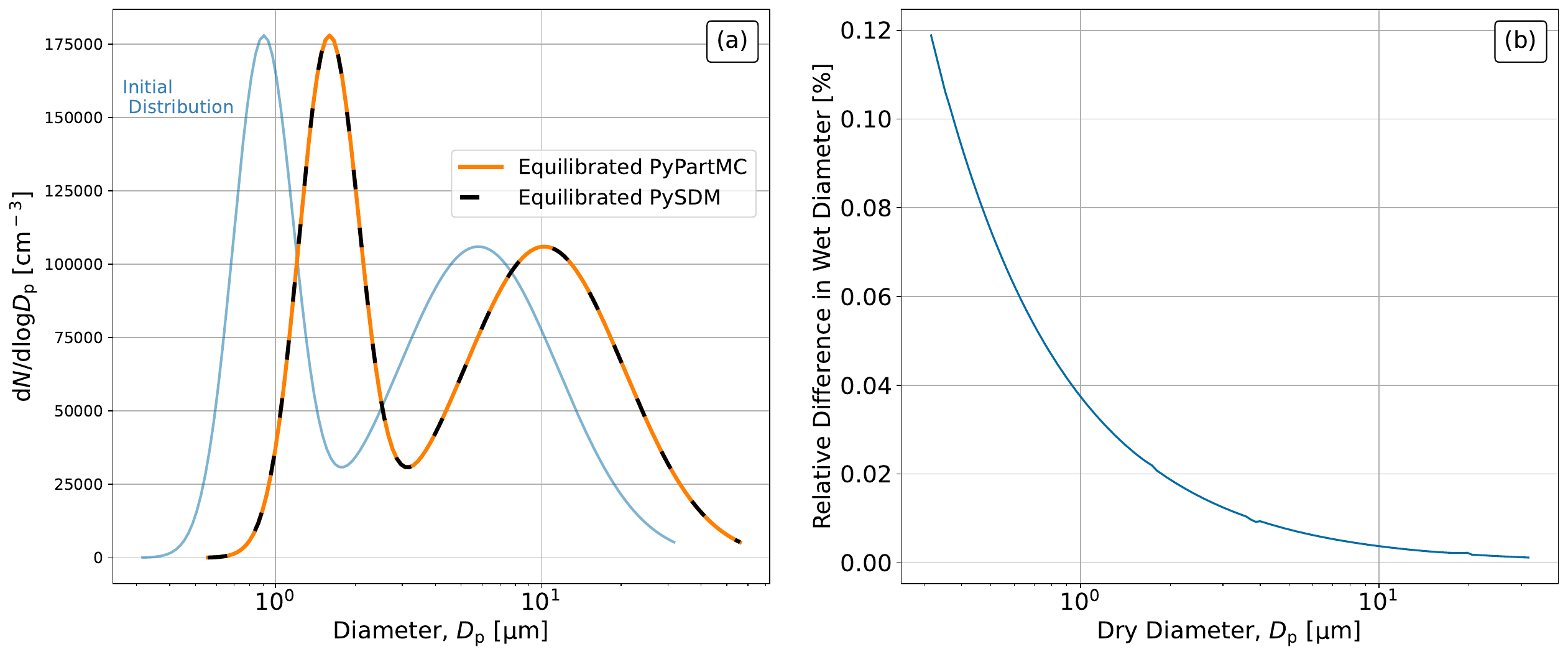}
\caption{\label{dry_wet_ex}
(a) Initial dry size distribution (light blue) and distributions after water uptake from PyPartMC and PySDM (orange and black, respectively) evaluated at a temperature of 295 Kelvin, relative humidity of 82\%, and hygroscopicity parameter kappa of 1.1. Note that the solutions from PyPartMC and PySDM are almost identical, and the orange and black curve are on top of each other. (b) Relative difference in wet diameter from PyPartMC and PySDM as a function of dry diameter.}
\end{figure}

This example compares PyPartMC directly to PySDM \citep{Bartman_2022}, another package for simulating particle dynamics. Computational particles are instantiated from a dry aerosol size distribution prescribed by interactive widget sliders. PyPartMC bindings pass information regarding the environmental state, such as relative humidity and temperature, size distribution parameters, and the hygroscopic parameter, kappa, to PartMC internals to evaluate equilibrium state under the given conditions. Shown in Figure~\ref{dry_wet_ex}(a), the initial dry distribution and the updated wet distribution predicted by PartMC are plotted alongside the result from PySDM. The two result curves from PartMC and PySDM look visually indistinguishable, however small numerical differences exist in calculating droplet size after the equilibriation with ambient relative humidity as shown in Figure~\ref{dry_wet_ex}(b). This is to be expected given that two different code bases were used. The use of a Jupyter notebook with Python kernel allows figures to be easily produced and downloaded using matplotlib \citep{matplotlib}. This particular example showcases the utility of employing PyPartMC to foster purposeful comparisons between different modeling frameworks, facilitated by both packages being exposed to Python and ready-to-use in a web-based notebook with interactive output. Additionally, PyPartMC has been incorporated into the PySDM v2 test suite to run automated checks against PartMC for this same equilibrium example \citep{deJong_et_al_2022}.

\subsection{Example \#2: Urban plume simulation}
In another example, we implemented a modified scenario based on \cite{Riemer_2009}
which considered an idealized urban plume. The scenario tracks the evolution of an
air parcel which is advected over an urban area. During the advection process, the parcel 
experiences emissions of gases and aerosols for 12 hours. After 12 hours of simulation,
the emissions are switched off and the parcel continues to evolve. The scenario presented here 
consists of gas and aerosol emissions, gas and aerosol dilution with the background, and particle
coagulation. This differs from the case presented in \cite{Riemer_2009} as
 gas phase chemistry, gas-aerosol partitioning and aerosol thermodynamics were not included. Normally, these components are incorporated by coupling the PartMC model with MOSAIC \citep{zaveri2008}. However, since MOSAIC code is only available upon request and not available as open source, we do not include it here.

\begin{figure}[ht]
\centering
\includegraphics[width=3.08in]{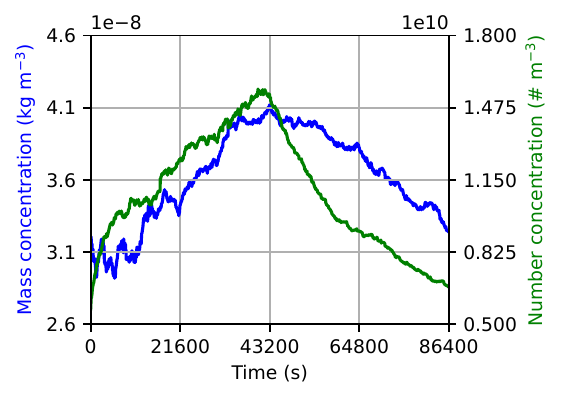}
\caption{\label{fig_urban_plume}
Evolution of total number concentration and total mass concentration.}
\end{figure}

In Figure~\ref{fig_urban_plume}, we show the time evolution of the total
aerosol number and mass concentration.
This figure is constructed by calling PyPartMC functions that sum up the number concentrations and the total mass concentration of each particle.

\begin{figure}[ht]
\centering
\includegraphics[width=\textwidth]{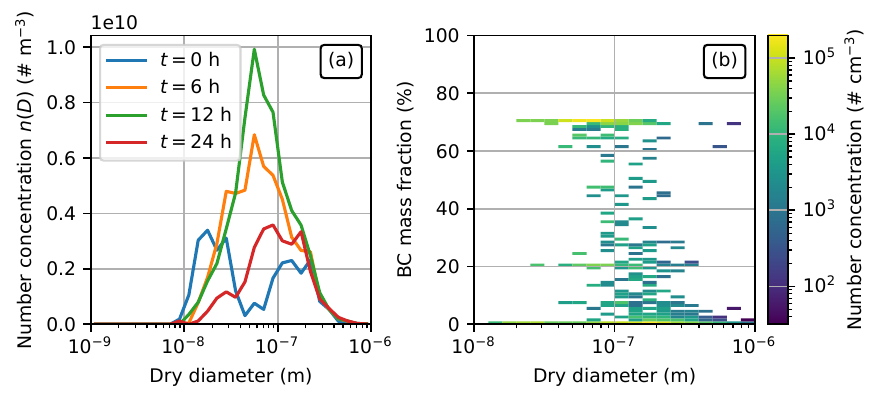}
\caption{\label{fig_urban_plume_dists}
(a) Number distributions $n(D)$ for the simulation with coagulation after 0, 6, 12 and 24 hours.
(b) Two-dimensional number distribution of black carbon dry mass fraction and particle dry diameter after 24 hours of simulation.}
\end{figure}

Figure~\ref{fig_urban_plume_dists}(a) shows the evolution of the initial aerosol size distribution after 6, 12 and 24 hours.
This figure is constructed by calling functions to compute particle dry diameters and acquire arrays of number concentrations
associated with each computational particle. 
With these two variables, a 1D histogram function is used to construct particle size distributions.
Figure~\ref{fig_urban_plume_dists}(b) shows the two-dimensional
distribution of black carbon mass fraction and particle dry diameter
after 24 h of simulation time.  Here the PyPartMC API is used to compute arrays of particle diameters
and also acquire black carbon mass, total dry mass and number concentration of each computational particle.
To analyze the data, a 2D histogram is created using arrays of dry diameters and black carbon mass fraction
with each computational particle weighted by its respective number concentration.

\subsection{Example \#3: Optical properties calculation using external python package}

\begin{figure}[h]
\centering
\includegraphics[width=3.08in]{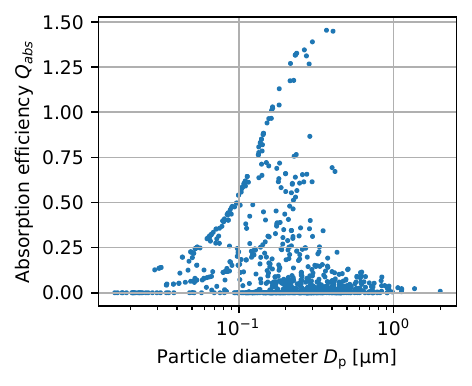}
\caption{\label{fig_optical}
Absorption efficiency of individual particles as a function particle diameter.
Each point represents an individual particle.}
\end{figure}

As a final example, we implemented a particle simulation where we
incorporated calls to an external library (PyMieScatt) that computes the optical properties of particle scattering and absorption \citep{Sumlin2018}.
PyPartMC evolves the aerosol population and then computes
the per-particle core diameters and total diameters.
Here Python controls the timestepping using a ``time block'', which passes to Fortran a time interval
consisting of many time steps to simulate. For this example, we chose a time block of 3600 seconds
and an internal model time step of 60 seconds.
Upon completion of each time block, a function to compute single-particle optics
is called by passing the AeroState type, which is a collection of AeroParticles.
Inside this function, arrays of particle core diameter and particle total diameter are constructed,
and the entire list of particles is iterated over to compute the scattering and absorption efficiency of each particle.
Additionally, the function computes bulk optical coefficients of scattering and absorption
by computing each particle's cross sectional area multiplied by optical efficiencies and weighted
by each computational particle's number concentration.
Overall, the involved arrays can be quite large, since they have the dimension of $N_{\rm particles} \times N_{\rm species}$ (on the order of 10,000 $\times$ 20), and having an API interface facilitates the analysis since it avoids exporting to files using the wide range of data formats supporten by Python libraries (including JSON, netCDF).
Figure~\ref{fig_optical} shows the results of computing absorption efficiencies of all particles
in the population at a given time.
This example shows how PyPartMC makes available a large quantity of
data in the Python environment that can be easily passed to other Python libraries.
Possibilities of external libraries include other optical models as well as
gas/aerosol chemistry and machine learning codes,
all of which may potentially feedback to the PyPartMC simulation as applicable.

\section{Impact}

Removing limitations to and streamlining the use of PartMC with PyPartMC will facilitate the dissemination of computational research results through independent execution of simulation runs based on PartMC, which could prove advantageous during the peer review process. Additionally, the ability to easily package examples, simple simulations, and results in a web-based notebook allows PyPartMC to support the efforts of many members of the scientific community, including researchers, instructors, and students, with nominal software and hardware requirements.

Researchers familiar with Python can take advantage of the PartMC simulation suite through the PyPartMC interface to tackle their own unique research questions and interests without ever needing to know about Fortran's involvement, while still gaining access to PartMC internals (Fortran derived types exposed as Python classes). This distinction makes the utility of the PartMC codebase more accessible and usable to those outside of the scientific modeling enclave. 

Python interfacing provides the added benefit of full access to NumPy~\citep{numerical_python}. NumPy's position as the foundational array programming library for Python has supported research endeavors in numerous fields, and its memory model makes passing and manipulating arrays between Python, C, and Fortran straightforward \citep{Harris_2020}. It also facilitates the integration of PartMC with python-based machine learning platforms such as PyTorch \citep{NEURIPS2019_9015} as well as with other programming languages through prominent Python-bridging tools (e.g., PyCall for the Julia language, see Fig.~\ref{fig:basic_examples}).

More frequent comparisons between experiments, measurements, and simulations to include PyPartMC guide further development and collaborative improvements. Visible development contributes to transparency in research and enables easy tracking of model performance over time. The open-source nature of PartMC in conjunction with PyPartMC's flexibility in external coupling empowers users to modify the code to suit their own unique research goals and tackle complex scientific questions with any number of interoperable software libraries and frameworks at their disposal. 

\section{Conclusions}

Open-source scientific projects strive to make research code more accessible to the wider scientific community by mitigating the steps and effort needed to apply these resources while simultaneously providing the infrastructure necessary to couple with existing tools. Increasing accessibility and usability of scientific software introduces more diversity to the community as researchers and students from various backgrounds and institutions can implement and advance the same project.

In this paper, we highlight the role of a Pythonic interface in making an existing open-source aerosol model more accessible and usable to those with less computational experience. Installation (or upgrade) of PyPartMC and all dependencies is reduced to a single command, permitting access to the functionalities of the PartMC codebase on Linux, macOS, and Windows. Example Jupyter notebooks exhibit how PyPartMC may be used in a Python environment to assist in executing meaningful comparisons between PartMC and other modeling frameworks, paving the way for more robust model benchmarking exercises. This capability also demonstrates the ease of replicating and disseminating research results with PyPartMC.

\section*{Code availability}

PyPartMC and all its dependencies are free/libre and open-source software.
Figures presented in the paper can be independently recreated using code packaged with PyPartMC.
Each release of the package is persistently archived at Zenodo: \url{https://doi.org/10.5281/zenodo.7662635}.

\section*{Author contributions}

NR and MW (lead authors of PartMC) defined the goals for the project, supervised the team and managed the funding.
PartMC architecture (Fig.~\ref{fig:architecture}), JSON input handling, compilation/testing/packaging automation and the Julia example were devised by SA.
API design and the wrapping code for exposed classes \& functions were developed  by JC and ZD with code reviews from SA.
Code presented in Fig.~\ref{fig:basic_examples} was developed by ZD and SA.
Example notebooks which are part of PyPartMC package have been developed by ZD (incl. Fig.~\ref{dry_wet_ex}) and JC (including Figures~\ref{fig_urban_plume}--\ref{fig_optical}).
AV contributed to packaging infrastructure and to web-app deployment setup using Voilà. 
Maintenance of PyPartMC has been carried out by SA, ZD, and JC.
The~paper was written by ZD with further feedback from all authors.

\section*{Conflict of Interest}

We wish to confirm that there are no known conflicts of interest associated with this publication and there has been no significant financial support for this work that could have influenced its outcome.

\section*{Acknowledgements}
This study was supported by the U.S. Department of Energy’s Atmospheric System Research, an Office of Science Biological and Environmental Research program, under grants DE-SC0021034 and DE‐SC0022130, and by the National Science Foundation under grant NSF-AGS 19-41110.
SA acknowledges support from the Polish
National Science Centre (grant no. 2020/39/D/ST10/01220).





\bibliographystyle{elsarticle-num}
\bibliography{refs_pypartmc}

\end{document}